\documentclass[12pt]{article}
\usepackage{graphics}
\usepackage{epsfig}
\usepackage{epstopdf}
\usepackage{amsmath}
\usepackage{array}
\usepackage[numbers,sort&compress]{natbib}
\begin{document}
\begin{center}
{\bf \Large {
	Contrasting the bulk viscous model with the standard $\Lambda$CDM using Bayesian statistics}} \\[0.2in]
V Mohammed Shareef, ND Jerin Mohan and Titus K Mathew\\
Department of Physics, Cochin University of Science And Technology, Kochi 22, India\\\

{shareef.morayur@cusat.ac.in\\ jerinmohandk@cusat.ac.in
\\ titus@cusat.ac.in} \\[0.3in]
\begin{abstract}
	Causal dissipative model is a plausible choice to explain the late accelerated epoch of universe. At the same time $\Lambda$CDM is considered to be standard model to explain the cosmological data corresponding to the late evolution of the universe. We consider a bulk viscus model in which the dissipation is driven by the bulk viscosity $\zeta=\alpha \rho^{1/2}$, described using the full causal Israel-Stewart theory. We computed the model parameters using the latest observational data (Pantheon). We contrasted this model with $\Lambda$CDM model for the late accelerated phase using Bayesian inference method. The Bayes factor has been obtained by calculating the likelihood for Pantheon data. Suitable prior values has been assumed for model parameters for calculating the likelihood. It shows that the evidence for $\Lambda$CDM against this viscous model is very strong according to Jeffreys scale.
\end{abstract}
\end{center}
\section{Introduction}
   The current understanding of the universe is that it is accelerating in expansion during the present epoch \cite{Reiss,Perlmutter,Bennet,Tegmark,Seljak,Komatsu}. This late acceleration can be attributed to the presence of a dominant exotic cosmic component called dark energy. The standard $\Lambda$CDM model, the most successful model for explaining this accelerated expansion, incorporates the cosmological constant as dark energy. However, the model is plagued mainly with the cosmological constant problem and the coincidence problem. The huge discrepancy between the observed and predicted values of the cosmological constant density, which is of the order of 120, is the cosmological constant problem. The coincidence between the energy densities of the dark energy and dark matter components during the current epoch of the universe is adjudged as the coincidence problem. 
 Further the $\Lambda$CDM model plagued with significant disagreement between the current value of the Hubble parameter measured by the classical distance ladder and that of the Planck CMB data \cite{WL,Wendy}. To be specific, we have $H_0=74.03 \pm 1.42$ km/s/Mpc from the Supernovae observation \cite{AG}, while the $\Lambda$CDM cosmology deduced from Planck CMB data favours  $H_0 = 67.4 \pm 0.5$ km/s/Mpc \cite{Aghanim}. Various dynamical dark energy models have been proposed to alleviate these issues 
\cite{Weinberg,Fujii,Carroll,Ford,Copeland,Chiba,Armendariz,EJ}. Quintessence dark energy \cite{JK}, ghost dark energy model \cite{Veneziano,Rosenzweig}, holographic model \cite{Thomas,Praseetha}, k-essence \cite{Picon}, phantom \cite{RR}, tachyon \cite{T}, dilaton \cite{M}, quintom \cite{E} and dynamical vacuum energy \cite{Gomez} are few examples of the dynamical DE models. Many of them achieve good consistency with observations. Modification of gravity theory is another alternative approach
\cite{Capozziello,Sotiriou,Nojiri, Ferraro,Myrzakulov, SD, Padmanabhan, Horava, Amendola, Dvali}.

Recently, dissipative models with bulk viscosity has been an intense field of research, since its ability to predict the late accelerated expansion by a unified description of the dark sectors. It is the bulk viscosity associated with the matter sector, that is shown to cause the late acceleration of the Universe.  Bulk viscous pressure can be originated whenever a system gets deviated from its thermal equilibrium, and is manifested as the one that brings it back to its original equilibrium state. There are two formalism in use to account for the bulk viscosity in cosmological theories viz. Eckart theory developed by Eckart \cite{Eckart} and others, and Israel-Stewart (IS) theory \cite{Israel,Stewart}.  Eckart theory is non causal, in which the perturbations propagate with infinite speed and the final equilibrium states achieved are unstable. It takes account of only the first order deviation form equilibrium and as a result, the relation between viscous pressure and rate of expansion of the Universe is linear \cite{Coley}. Hence, it is relatively easy to analyse the evolution of various cosmological parameters. On the other hand, IS theory takes account of the higher order deviations from equilibrium in the dissipative fluids and hence the theory is causal, and the perturbations propagate only with finite speed. Moreover, the final equilibrium state obtained in this theory is stable \cite{Hiscock}.  
The IS theory has been used to describe the inflation during early epoch of evolution of the Universe, by Maartens and Mendez \cite{Maartens}, and they found that the results are thermodynamically consistent. Various other studies have also been performed using the theory \cite{Mak,Harko,Zimdahl,Zakari,Cooley}.

Cataldo et al. have studied late acceleration using IS formalism where the authors have used an ansatz for the Hubble parameter and have shown that the universe might have undergone a transition to the phantom behaviour leading to big-rip singularity \cite{Cataldo}. Piattella et al. have tried to unify dark matter and dark energy, found numerical solutions to the gravitational potential using an ansatz for the viscous pressure depends on the density of the viscous fluid and compared it with the standard $\Lambda$CDM model \cite{Piattella}. They concluded that, in gross comparison with the standard $\Lambda$CDM model, the viscous model with the full causal theory leads to some disfavoured features compared to the truncated version of the model. So, by and large in solving the viscous model using the causal theory, it seems that many have used some ansatz either for the Hubble parameter or for the viscous pressure.

A recent use of the full IS theory to study the recent acceleration of the Universe, by invoking the bulk viscosity associated with the matter sector has been carried out in detail, with reference to the background evolution of the universe in \cite{Jerin}. The dynamical system analysis and thermal evolution studies of the same model have been done by the same authors in \cite{ND}. It has been shown that the model can predict the background evolution of the universe, which end with a de Sitter type epoch. Using the state-finder diagnostics, the model has been contrasted with the $\Lambda$CDM model and have shown that it is arguably different from the standard model. We think it is worth to contrast this model with the standard $\Lambda$CDM model using more stringent statistical tools. In this work, we aim to find the relative significance of the IS-bulk viscous model in explaining the recent acceleration of the universe, in comparison with the standard $\Lambda$CDM model. We are using the Bayesian analysis method for this comparison with the aid of latest data on Supernovae, the Pantheon data.

The paper is organized as follows. In Sect. 2, we review the bulk viscous model of the late accelerating universe. In Sect. 3, we extract the Hubble parameter and bulk viscous parameters of our model for Pantheon data. In Sect. 4, we perform the Bayesian analysis, and the conclusion is discussed in Sect. 5.

\section{The Bulk viscous model of the late accelerating universe}
In this section, we present a brief review of the model in concern. Jerin et al. have discussed this model in detail, in describing the background evolution of the universe \cite{Jerin,ND}. The authors began with the basic equations governing the evolution of universe
\begin{equation}
3H^{2}=\rho_{m}
\end{equation}
\begin{equation}
\dot{H}=-H^{2}-\frac{1}{6}(\rho_{m}+3P_{eff})
\end{equation}
where H=$\frac{\dot{a}}{a}$ is the Hubble parameter, $\rho_{m}$ is the matter density and
\begin{equation}
P_{eff} = p+\Pi
\end{equation}
is the effective pressure, p=($\gamma$-1)$\rho$ is the normal kinetic pressure with $\gamma$ as the barotropic index and $\Pi$ is the bulk viscous pressure. The equation governing the evolution of matter density is expressed as (including the bulk viscous pressure)
\begin{equation}
\dot{\rho_{m}}+3H(\rho_{m}+P_{eff})=0
\end{equation}
The expression for the evolution of bulk viscous pressure $\Pi$, according to the IS theory is given by
\begin{equation}
\tau\dot{\Pi}+\Pi = -3\zeta H-\frac{1}{2}\tau\Pi\left (3H+ \frac{\dot{\tau}}{\tau}-\frac{\dot{\zeta}}{\zeta}-\frac{\dot{T}}{T} \right )
\end{equation}
where are $\tau$, $\zeta$ and T are the relaxation time, bulk viscosity and temperature respectively, defined as
\begin{equation}
\tau=\alpha\rho^{s-1},\ \zeta=\alpha\rho^{s},\ T=\beta\rho^{r}
\end{equation}
Here, $\alpha$, $\beta$ and s are all positive constant parameters and r=$\frac{\gamma-1}{\gamma}$. For $\tau$ = 0, the eq. (5) reduces to the simple Eckart equation, $\Pi$ =-3$\zeta$H. The authors preferred $\gamma$=1 since it favours the late accelerated phase (dominated with non relativistic matter) and prefer s=1/2 since it possesses a prior decelerated epoch which is unstable and a stable future accelerated epoch \cite{ND}, then
\begin{equation}
\tau=\alpha\rho^{-1/2},\ \zeta=\alpha\rho^{1/2},\ T=\beta
\end{equation}
Eq. (1) combining with eqs. (3) and (4), an expression for the bulk viscous pressure $\Pi$ is obtained as
\begin{equation}
\Pi=-2\dot{H}-3H^{2}
\end{equation}
Substituting in eq. (5), combining with eq. (7), one gets
\begin{equation}
\ddot{H}+\left (\dfrac{3}{2}+\dfrac{\sqrt{3}}{\alpha}\right)H\dot{H}-\dfrac{\dot{H}^{2}}{H}+\left (\dfrac{3^{3/2}}{\alpha}-\dfrac{9}{4}\right)H^{3}=0
\end{equation}
The above equation admits the solution of the form
\begin{equation}
H=H_{0}\ (C_{1}a^{-m1}+C_{2}a^{-m2})
\end{equation}
where $H_{0}$ is the present value of Hubble parameter and the other constants are as in \cite{Jerin},
\begin{equation}
C_{1;2}=\dfrac{\pm 1+\sqrt{1+6\alpha^{2}}\mp\sqrt{3}\alpha\tilde{\Pi_{0}}}{2\sqrt{1+6\alpha^{2}}}
\end{equation}
\begin{equation}
m_{1;2}=\dfrac{\sqrt{3}}{2\alpha}\left(\sqrt{3\alpha}+1\pm\sqrt{1+6\alpha^{2}}\right)
\end{equation}
where $\tilde{\Pi_{0}}$=$\dfrac{\Pi_{0}}{3H_{0}^{2}}$ is the bulk viscous pressure parameter (dimensionless), with $\Pi_{0}$ as the present value of $\Pi$.

The authors have studied the nature of evolution of scale factor a and deceleration parameter q in order to analyse whether the model is in agreement with the current understanding regarding the evolution of Universe or not. For this, the authors have constrained the model with cosmological observational data on Supernovae type Ia, the Union 2.1 data set consists of 307 data points. The studies based on the computed model parameters, have shown that the model reasonably explains the observed transition into the late acceleration of the universe. The transition redshift is estimated to be $z_{T}$ $\sim$ $0.52^{+0.010}_{-0.016}$ and the present value of deceleration parameter is $q_0 \sim -0.59^{+0.015}_{-0.016}$. These predicted values are in the range obtained by WMAP data analysis. These findings confirm that the Universe described by the causal viscous pressure evolves from an early decelerated phase to a late accelerating phase \cite{Jerin}.

The age of Universe calculated in this model is around 9.72 Gyr and is considerably less than the conventional  observational results. The present value of the equation of state parameter, $\omega_{0}$, is found to be $\omega_{0}$ $\sim$ $-0.73^{+0.01}_{-0.01}$ and is evolving to an asymptotic value of -0.79 in the future phase, indicates a transition from stiff nature to quintessence nature. Statefinder studies of the model gives the value ($r_{0},s_{0}$) = (0.582, 0.128), indicating that the model is obviously distinct from that of the standard $\Lambda$CDM model. Also, the bulk viscous parameter satisfies $\zeta$ $\geq$ 0 evolves from a large positive value in the beginning and end up with a smaller value in the future. This indicates that the entropy production is always positive and the Local Second Law (LSL) is satisfied \cite{Jerin}.

The team has studied the dynamical and thermal behaviour of the model in their next paper \cite{ND} in detail. Assuming bulk viscosity to be $\zeta=\alpha \rho^{s}$, two different cases s= 1/2 and s$\neq$1/2 have been studied. The choice s=1/2 allows a stable evolution of Universe from early decelerated phase to late accelerated phase. The expansion is tending towards a state of maximum entropy and the Generalised Second Law (GSL) is valid. The choice s$<$1/2 results in a static universe and thus it is discarded. The choice s$>$1/2 gives a de Sitter universe (unstable), where the GSL is satisfied with an unstable thermodynamic evolution similar to the dynamical system behaviour.

\section{Data analysis for model parameters}
 Here, we evaluate the best-fit model parameters for $ \alpha $, $\tilde{\Pi_{0}}$ and the present value of Hubble parameter $ H_{0} $ by constraining with Pantheon data, having 1048 data points. The parameter values are obtained using the $\chi^{2}$ minimisation method.

In a flat universe, the luminosity distance $d_L$ is expressed as

\begin{equation}
	d_L(z,\alpha, \tilde{\Pi_{0}}, H_{0})= c(1+z)\int_{0}^{z} \dfrac{dz'}{H(z',\alpha, \tilde{\Pi_{0}}, H_{0})}
\end{equation}
The difference between apparent and absolute magnitude of luminosity distance depends on the $d_L$. The equation that relates
the theoretical distance moduli $\mu_{th}$, the apparent magnitude
m, the absolute magnitude M and $d_L$ is given by
\begin{equation}
	\mu_{th}(z,\alpha, \tilde{\Pi_{0}}, H_{0})= m-M= 5 \ log_{10} \left [\dfrac{d_L(z,\alpha, \tilde{\Pi_{0}},H_{0})}{Mpc}\right ] +25
\end{equation}
The observational distance modulus $\mu '_{i}$ obtained from
\begin{figure}
	\centering
	\includegraphics[scale=0.35]{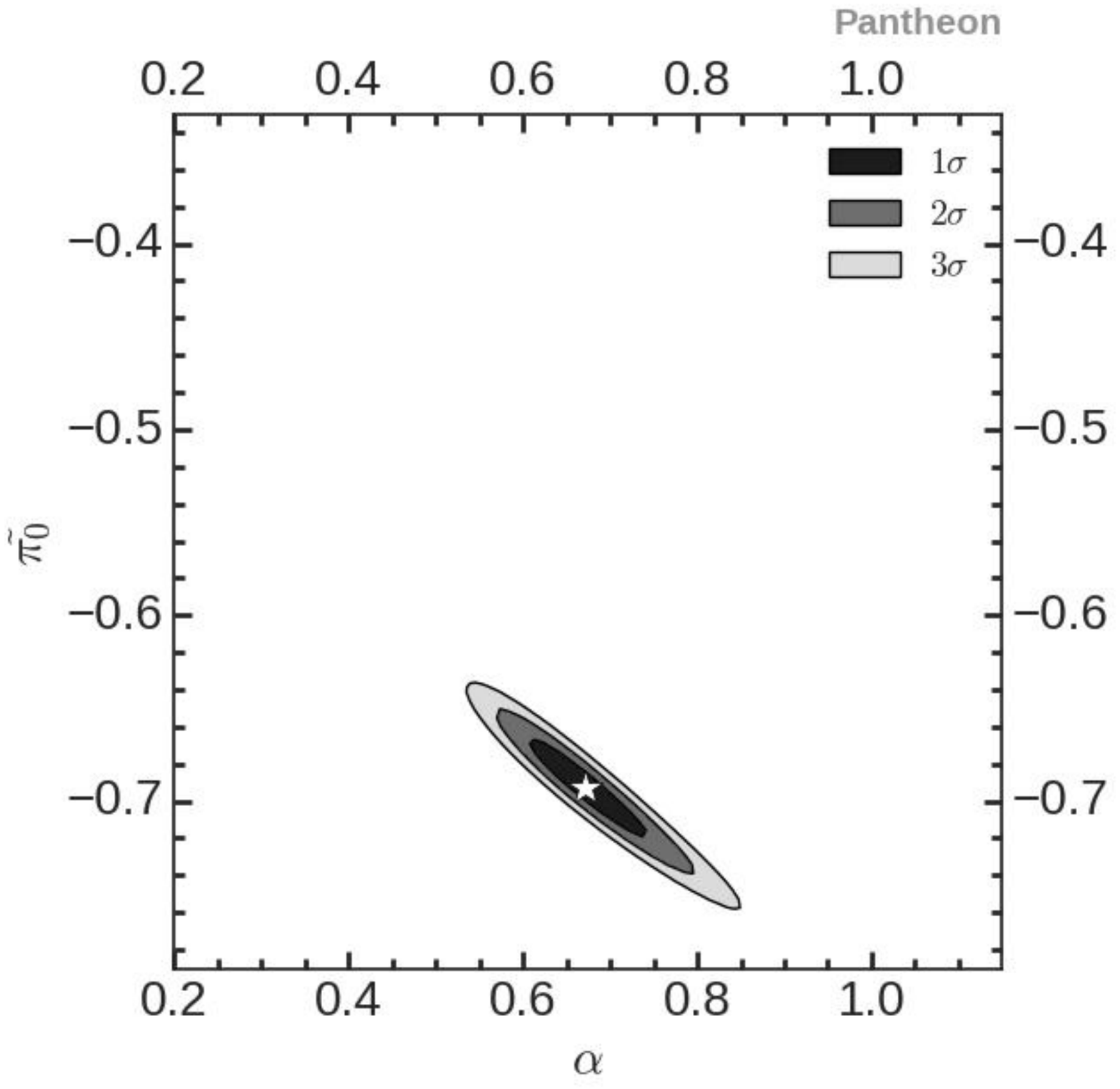}
	\caption{The confidence intervals for the model parameters $\alpha$ and $\tilde{\Pi_{0}}$  corresponding to 68.3, 95.4 and 99.73$\%$ probabilities. The best estimated values of the model parameters is indicated by a star.}
\end{figure}
Pantheon data set is compared with $\mu_{th}$ calculated using eq. (14), corresponding to different values of the redshifts. The $\chi^{2}$ function can be expressed as
\begin{equation}
	\chi^{2}(\alpha, \tilde{\Pi_{0}}, H_{0})= \sum_{i=1}^{n} \dfrac{\left [\mu_{th}(z,\alpha, \tilde{\Pi_{0}},H_{0})-\mu '_{i}\right ]^{2}}{\sigma_{i}^{2}}
\end{equation}
where n is the total number of data points and $\sigma_{i}^{2}$ is the variance of the $i^{th}$ measurement. The best estimated values of the parameters $\alpha$, $\tilde{\Pi_{0}}$, and $H_{0}$ have been obtained by $\chi^{2}$ minimization. For comparison, the parameter values for the $\Lambda$CDM model have also been extracted using the same data. The best estimated parameter values are shown in Table 1. The
$\chi^{2}_{min}$ function per degree of freedom is defined as  $\chi^{2}_{d.o.f}=\chi^{2}_{min}/(n-n^{\prime})$, where $n^{\prime}$ is the number of parameters in the model. Here n=1048 and n'= 3. The present value of the Hubble parameter obtained from the CVM model is comparable with that of the $\Lambda$CDM model. In reference \cite{Jerin}, parameters of the similar model were extracted using the Union Supernovae data sample consists of only 307 date points. But, we have used the latest data of Supernovae observation, the Pantheon sample, which consists of 1048 data points. The computation using the latest data are given in table \ref{Table:1}, gives parameter values which are different from the earlier prediction. The values of $\alpha, \Omega_{m0}, z_{T}$ and $H_0$ decreases slightly, compared to their earlier values, $\alpha= 0.66, H_{0} = 70.29, \Omega_{m0}= 0.283, z_{T} = 0.38.$ But in the case of other parameters, like equation of state, deceleration parameter, etc. there is a slight increase, compared to the previously evaluated values, $\omega_{0} = -0.73, q_{0}= -0.54$. The confidence interval plane for the model parameters $\alpha$, and $\tilde{\Pi}_{0}, $ corresponds to 68.3, 95.4 and 99.73$\%$ probabilities is shown in Fig. 1.
The probabilistic correction to the parameter values corresponding to 68.3$\%$ probability are also included in Table 1.
\begin{table}
	\centering
	\begin{tabular}{ |p{1.25cm}|p{1.4cm}|p{1.8cm}|p{0.9cm}|p{2.5cm}|p{1.25cm}|p{0.85cm}|}
		\hline
		\vspace{0.1cm}&\multicolumn{6}{|c|}{Best estimated values of parameters} \\[8pt]
		\hline
		\vspace{0.1cm}Model&\vspace{0.1cm}$ \alpha $&\vspace{0.1cm}$  \tilde{\Pi}_{0}$&\vspace{0.1cm}$ \Omega_{m0} $&\vspace{0.1cm}$ H_{0} $\begin{small}
			($ kms^{-1}Mpc^{-1} $)
		\end{small}&\vspace{0.1cm}$ \chi^{2}_{min} $&\vspace{0.1cm}$ \chi^{2}_{d.o.f.} $\\ [8pt] 
		\hline
		\vspace{0.1cm}
		CVM model &\vspace{0.1cm}$0.67^{+0.07}_{-0.07}$ &\vspace{0.1cm}$-0.69^{+0.03}_{-0.03}$&\vspace{0.1cm}1& \vspace{0.1cm}70.35& \vspace{0.1cm}1034.50&\vspace{0.1cm}0.980 \\[8pt]
		\hline 
		\vspace{0.07cm}$\Lambda$CDM model&\vspace{0.07cm} \ - &\vspace{0.07cm} \ - &\vspace{0.07cm} 0.283& \vspace{0.07cm}70.25& \vspace{0.07cm} 1035.71 &\vspace{0.07cm} 0.991\\[8pt]
		\hline
	\end{tabular}
	\caption{A comparison of best estimated values of CVM model parameters with the standard $\Lambda$CDM model for Pantheon data}
	\label{Table:1} 
\end{table}

\section{Bayesian analysis for the model}
In this section, we describe the final part of our analysis, that is obtaining the Bayes factor is obtaining the significance the bulk viscous model in explaining the late acceleration of the universe using the full IS theory.
It has long been recognised by the scientific community that Bayesian analyses can give reliable results on model selection
\cite{Trotta}.  Cosmological model selection refers to comparing different models based on their reliability to describe the observational data. Basically, Bayesian analysis depends both on the data properties and the prior parameter ranges of the models being compared. Bayesian comparison is based on Bayes theorem, given by a simple relation: 
 \begin{equation}
  P(M | D, I) = \frac{P(D|M,I) P(M|I)}{P(D|I)}
 \end{equation}
where $P(M|D,I)$ is the posterior probability of the model $M$ given the data set $D$ and $I$ the background information are true, $P(D|M,I)$ is the probability of the data given the model and the background information are true, also known as the likelihood of the model, $P(M|I)$ is known as the prior probability of the model assigning on the basis of the background information and $P(D|I)$ is the probability for obtaining the data provided the background information is true and is of the status of a normalisation factor.

Using the above equation, we obtain the posterior probability of a model \cite{Liddle,AR,Davis,Kurek,Mortlock,Santos,John}, and in our case, it is the causal bulk viscous cosmological model, $M_{CVM},$ based on IS theory. Let $P(M_{CVM}| D, I)$ be the posterior probability of $M_{CVM}$, such that $D,$ the given set of observational data and $I,$ the background information regarding the expansion of the universe are true \cite{Taylor}. Then eq. (16) can be expressed as. 
\begin{equation}
P(M_{CVM}|D, I)=\dfrac{P(M_{CVM}|I)\ P(D|M_{CVM}, I)}{P(D|I)}
\end{equation}
where P($M_{CVM}|$I) is the prior probability assigned to the model before data analysis, given I is true, P(D$|M_{CVM}$, I) is the probability to obtain the data for which the respective model and I are true, called the likelihood, L($M_{CVM}$) of the model, , and $P(D|I)$ is the probability of obtaining the data provided $I$ is true. That is {\it the likelihood updates the prior model probability to the posterior model probability, 
i.e. the probability of the model given the data.} The same method can be performed in order to deduce the likelihood of $\Lambda$CDM model too, L($M_{\Lambda CDM}$), for the given set of data. Now we define a parameter $O_{ij}$ called odds ratio, the ratio of posterior probabilities of the models
\begin{equation}
O_{ij}=\dfrac{p(M_{CVM}|D, I)}{p(M_{\Lambda CDM}|D, I)} = \dfrac{p(M_{CVM}|I)\ p(D|M_{CVM}, I)}{p(M_{\Lambda CDM}|I)\ p(D|M_{\Lambda CDM}, I)}
\end{equation}
where $i$ and $j$ represent $M_{CVM}$ and $M_{\Lambda CDM}$ respectively, assuming the normalisation factor $p(D|I)$ be same for both models. The prior probability depends only on the previous knowledge of a person. If his knowledge does not prefer one model over the other, the prior probabilities get cancelled in the
odds ratio ($O_{ij}$). Then we have
\begin{equation}
O_{ij}= \dfrac{p(D|M_{CVM}, I)}{p(D|M_{\Lambda CDM}, I)}\equiv B_{ij}
\end{equation}
Here $B_{ij}$ represents the Bayes factor. 

By convention, the likelihood of the CVM model can be denoted as $L(M_i)$ where i denotes CVM. Taking $\alpha$, $\beta$ as the model parameters, likelihood can be expressed as \cite{John}
\begin{equation}
L(M_i)=\int p(\alpha|M_i) \ d\alpha \ \left( \int p(\beta|M_i) \ L(\alpha, \beta) \ d\beta \right)
\end{equation}
$L(\alpha, \beta)$ represents the likelihood of the combination of the model parameters. It can be deduced by marginalizing all the model parameters. Here we assume the errors of measurement form a Gaussian. Then the likelihood function can be expressed as \cite{Li}
\begin{equation}
L(\alpha, \beta)=e^{-\frac{\chi ^{2} (\alpha, \beta)}{2}}
\end{equation}
\begin{equation}
\chi ^{2} (\alpha, \beta)=\Sigma \left[ \dfrac{A_k - A_k (\alpha, \beta)}{\sigma_k} \right ] ^{2}
\end{equation}
where $A_k$ is the observed value, $A_k (\alpha, \beta)$ is the corresponding theoretical value and $\sigma_k$ is the uncertainty in the measurement of the observable. For the parameters $\alpha$ and $\beta$, we assume a uniform prior information such that they lie in the range $[\alpha,\ \alpha+\Delta \alpha]$ and $[\beta,\ \beta+\Delta \beta]$ respectively. As the simplest choice, the prior probability of parameters can be expressed as $p(\alpha|M_i)=\dfrac{1}{\Delta \alpha}$ and $p(\beta|M_i)=\dfrac{1}{\Delta \beta}$ . Now, eq. (20) can be written as
\begin{equation}
L(M_i)=\dfrac{1}{\Delta \alpha} \ \dfrac{1}{\Delta \beta}\int_{\alpha}^{\alpha+\Delta \alpha} d\alpha ' \int_{\beta}^{\beta+\Delta \beta} exp \left [-\chi ^{2} (\alpha ', \beta ')/2 \right ] \ d\beta '
\end{equation}

The term $\Delta$ indicates the prior range of the respective model parameters, which is the fundamental requirement to estimate the likelihood of a model, to be chosen without considering the data. Now, we discuss how we have chosen the prior values for the model parameters $\Omega_{m0}$, $H_0$, $\alpha$ and $\tilde{\Pi_{0}}$

(i) parameter $\Omega_{m0}:$ \ We have chosen the prior of $\Omega_{m0}$ to be 0$\leq\Omega_{m0}\leq$1. This range has been chosen in comparison with similar ranges that used in many references, for instance: 0$\leq\Omega_{m0}\leq$1 \cite{Weller,Gupta}, 0.2$\leq\Omega_{m0}\leq$1 \cite{Lima}, 0.01$\leq\Omega_{m0}\leq$1 \cite{Ryan} etc.

(ii) parameter $H_0:$ \ For the present value of Hubble parameter, we adopted a range, 65–75 km $s^{-1}Mpc^{-1}$. The recent Planck observations \cite{Aghanim} measured a value of $H_0$ as 67.4 ± 0.5 km$s^{-1}Mpc^{-1}$ . On the other hand, a higher value 73.24 ± 1.74 km$s^{-1}Mpc^{-1}$ was obtained for $H_{0}$ from the local expansion rate estimation \cite{Reiss.}. The authors in the reference \cite{Tripathi} consider a prior range 65–75 km$s^{-1}Mpc^{-1}$ that includes almost all  the previous calculated values for $H_{0}$, which is our selected prior range.

(iii) parameter $\alpha$ : \ The equation of state parameter $\omega$ can be expressed in a simpler form as $\omega^{\pm}$=$(1/\sqrt{3}\alpha)\left(1\pm\sqrt{1+6\alpha^{2}}\right) $\cite{Jerin}, where $\omega^{-}<$ -1/3 corresponds to the late accelerated epoch. The authors have been shown that $\alpha$ can be varied only as 0.2$\leq\alpha\leq$1.15 for the late phase. This is our preferred prior range.

(iv) parameter $\tilde{\Pi_{0}}:$ \ The deceleration parameter for the current epoch can be expressed as $q_{0}=(1/2)(1+3\tilde{\Pi_{0}})$ \cite{ND}. Using this expression, the authors have been shown that the range of $\tilde{\Pi_{0}}$ for the late accelerated phase will be constrained as -0.79 $\leq\tilde{\Pi_{0}}\leq$ -0.33. This is our preferred prior range.

Substituting the above prior values, computations have been performed for Pantheon data of 1048 data points, to deduce the likelihood of both models. We obtained the values as $1.309*10^{-228}$ and $5.639*10^{-229}$ for $\Lambda$CDM and CVM model respectively.

The essence of likelihood calculation is exposed by estimating and plotting the marginal likelihood of various model parameters. It is similar to the likelihood calculation except we treat the parameter in question as a constant. For instance, the marginal likelihood of the parameter $\alpha$ of the CVM model can be expressed as
\begin{equation}
L_{CVM}(\alpha)=\dfrac{1}{\Delta H_0}\dfrac{1}{\Delta \tilde{\Pi}_{0}} \int_{H_{0}}^{H_{0}+\Delta H_{0}} dH_0 \int_{\tilde{\Pi}_{0}}^{\tilde{\Pi}_{0}+\Delta \tilde{\Pi}_{0}} exp \left [-\chi ^{2} (H_{0},\alpha, \tilde{\Pi}_{0})/2 \right ]\ d\tilde{\Pi}_{0}
\end{equation}
Here, we integrate the likelihood function $e^{-\chi ^{2}/2}$  over $H_0$ and $\tilde{\Pi_{0}}$ for the prior range of each parameter, treating $\alpha$ as constant. Then, the marginal likelihood expression $L_{CVM}(\alpha)$ is obtained. The behaviour of $\alpha$ in the likelihood calculation is 
\begin{figure}
	\centering
	\includegraphics[scale=0.4]{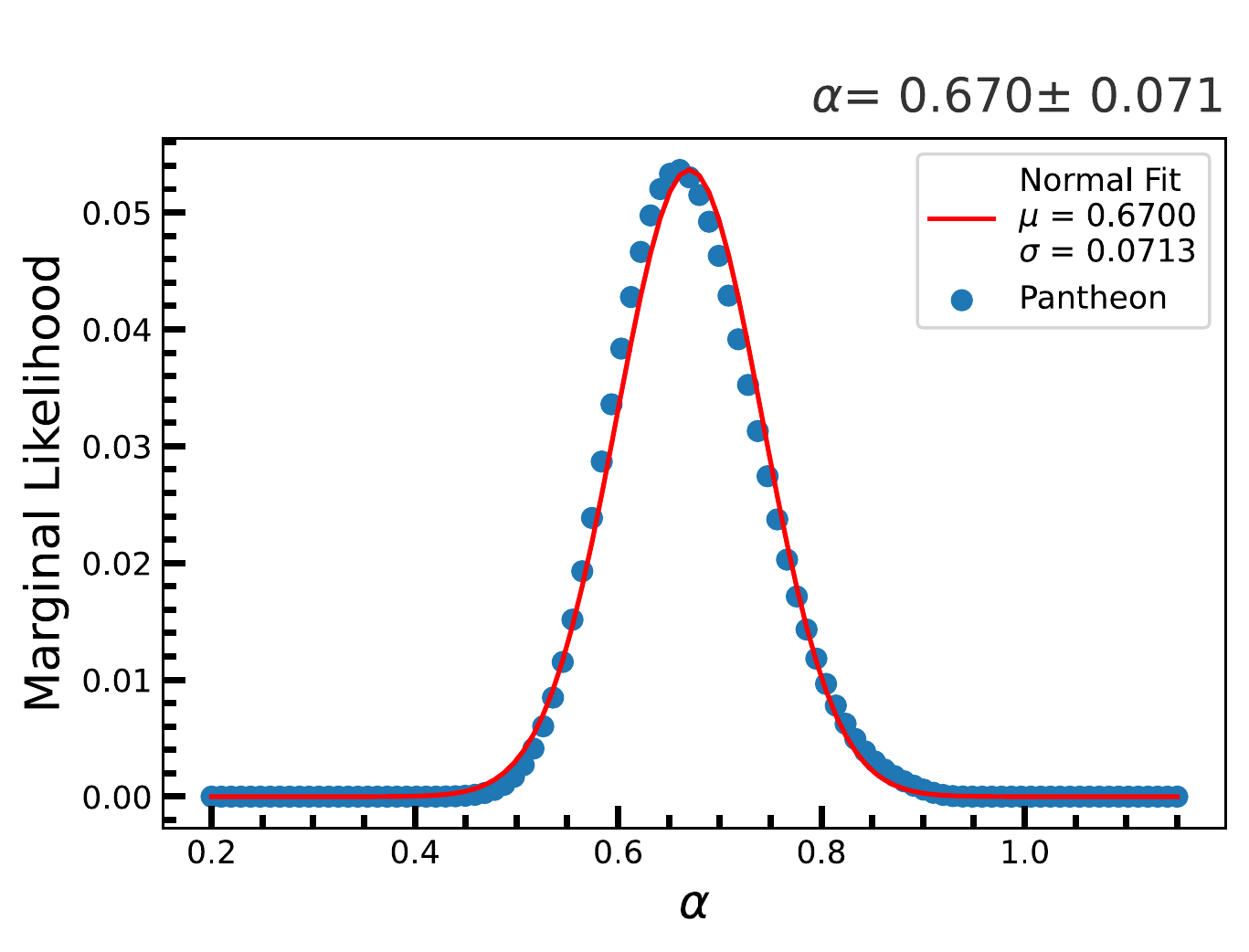}
	\includegraphics[scale=0.4]{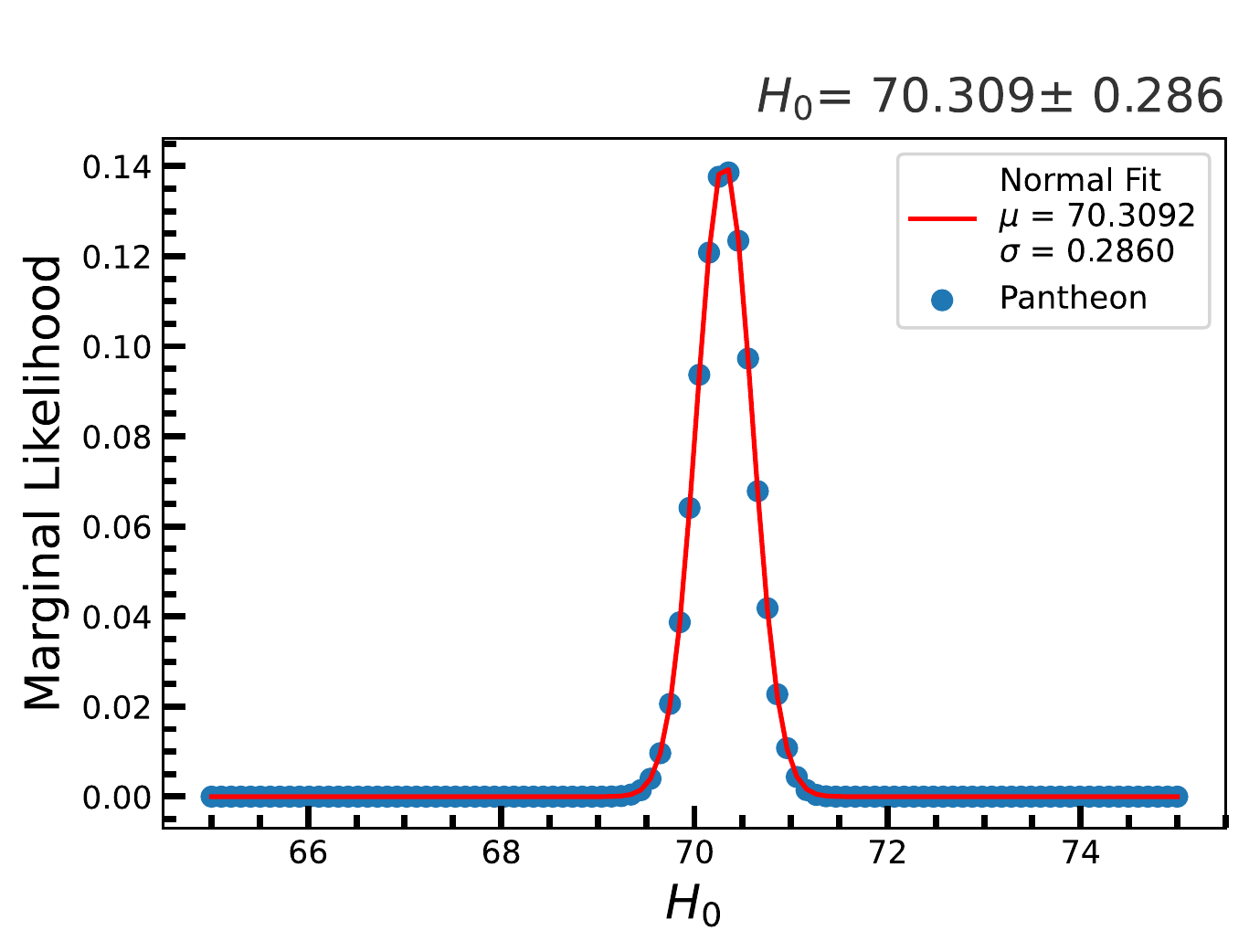}  
	\includegraphics[scale=0.4]{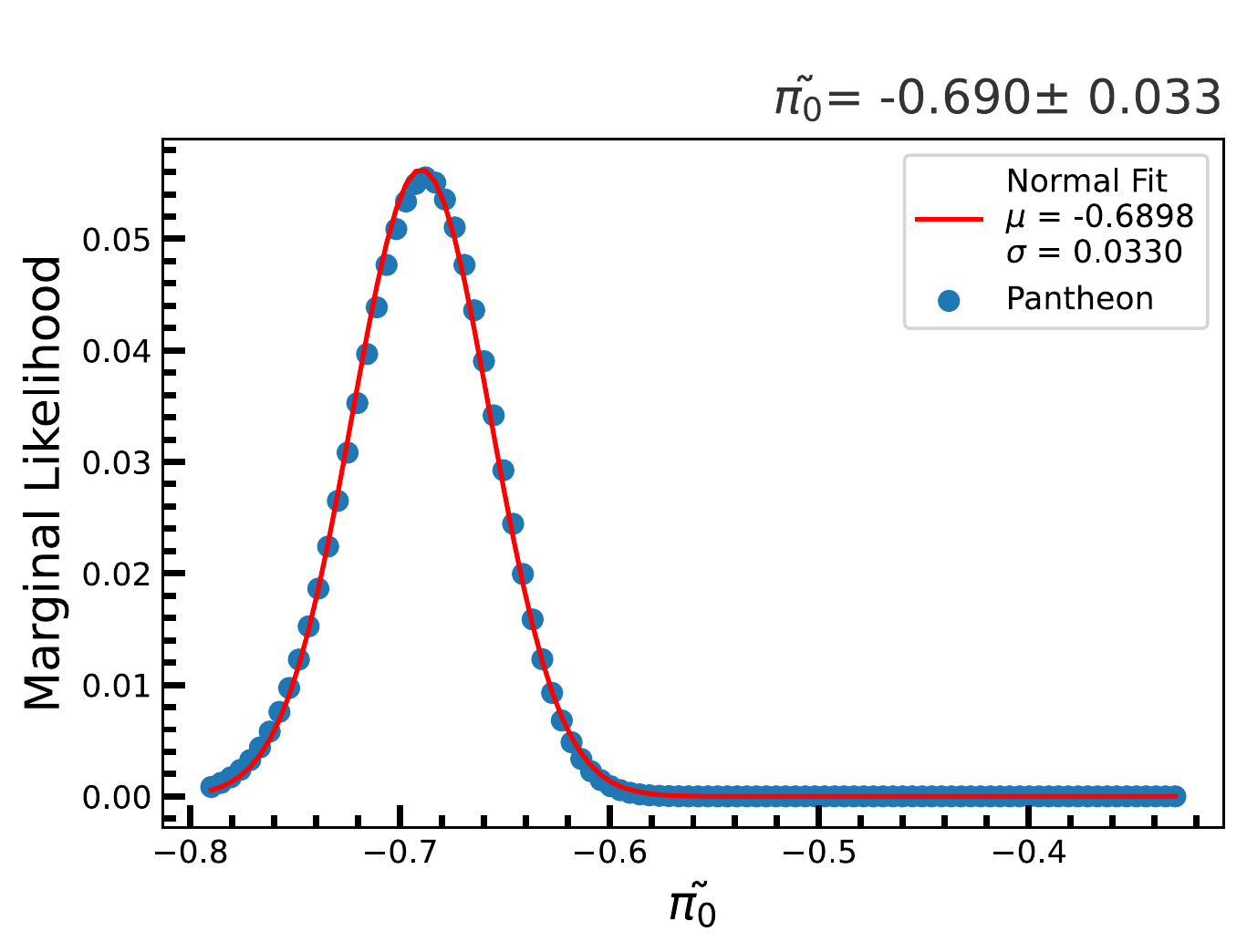}  
	\caption{Marginal likelihood of the model parameter $\alpha, H_0$ and $\tilde{\Pi}_0$ for Pantheon data}
\end{figure}
  exposed by plotting normalised $L_{CVM}(\alpha)$ against $\alpha$ (Fig.2). We obtained a true Gaussian curve peaked at the best estmated value of $\alpha$ obtained in the previous section using $\chi ^{2}$ minimisation technique ($\alpha$=0.67), with $\sigma = \pm 0.071$. The red line represents the curve and the blue dots represent the fitted Pantheon data points.
  \begin{table}
  	\centering
  	\begin{tabular}{ |p{6cm}|p{6cm}|}
  		\hline
  		\vspace{0.1cm}Model&\vspace{0.1cm} Likelihood \\ [3pt]
  		\hline
  		\vspace{0.1cm} CVM model &\vspace{0.1cm} $5.639*10^{-229}$ \\[3pt]
  		\hline
  		\vspace{0.1cm}$\Lambda$CDM model&\vspace{0.1cm} $1.309*10^{-228}$ \\[3pt]
  		\hline
  	\end{tabular}
  	\caption{Likelihood of CVM model and standard $\Lambda$CDM model for Pantheon data}
  	\label{Table:2} 
  \end{table}
In a similar fashion, the marginal likelihood of both $H_0$ and $\tilde{\Pi}_{0}$ are plotted (Fig. 3-4) to obtain true Gaussian curves and the Pantheon data is fitted with the curve. The variations of parameters are plotted for the prior range of each parameters. The present value of the Hubble parameter $H_0$ is peaked at $H_0$=70.309 is very similar to the value obtained using $\chi ^{2}$ minimisation 
with $\sigma = \pm 0.286$. The bulk viscous pressure parameter $\tilde{\Pi}_{0}$ peaked at exactly the same value obtained in the previous section ($\tilde{\Pi}_{0}$=-0.69), with $\sigma = \pm 0.033$.
 
 Fianlly, from eq. (19), the Bayes factor $B_{ij}$ can be estimated by taking the ratio of the likelihood of the two models, $M_i$ and $M_j$
 \begin{equation}
 	B_{ij}=\dfrac{L(M_i)}{L(M_j)}
 \end{equation}
The Bayes factor of model  corresponding to  the data sets are then calculated by comparing the likelihood of model and that of the standard $\Lambda$CDM model. The likelihoods are tabulated in table \ref{Table:2}
 Taking the ratio of the likelihood of both models, the Bayes factor is found to be around 0.431. As per the Jeffreys scale \cite{Jeff}, for $B_{ij}<1$,  $M_i$ is not significant with respect to $M_j$. For $1<B_{ij}<3$, there is evidence against $M_j$ when compared with, but not worth more than a mention. For $M_i$ $3<B_{ij}<20$, the evidence of  $M_i$ against $M_j$ is not strong, but definite. For $20<B_{ij}<150$ the evidence is strong and for $B_{ij}>150$, the evidence aginst $M_j$ is very strong \cite{Drell,Trotta, Athira}
 Since the Bayes factor is 0.431 which is less than 1, we can conclude that evidence of the bulk viscous model is very weak against the standard $\Lambda$CDM. It implies that, compared to the CVM model, the preference for the standard $\Lambda$CDM model is very strong.
 
\section{Conclusion}
Bulk viscous model of the universe, based on the full Israel-Stewart theory of causal viscosity, is considered to be a good alternative to explain the late accelerated epoch of the universe. In the present work, we have contrasted the model using the latest data on Supernovae. We have analysed the significance of this model in contrast to the standard $\Lambda$CDM model using the method of Bayesian analysis. 

We have extracted the model parameters using the Pantheon data and found that they are slightly different from the previously extracted values using an old dataset SN307. Following this, we proceed to contrast the viscous model with the standard $\Lambda$CDM model using the Bayesian statistics. 
We have obtained the likelihood of CVM using the extracted model parameters and also obtained the likelihood $\Lambda$CDM models using the same dataset. The basic ingredients for the likelihood are prior ranges for the parameters of models. We have used suitable flat priors for different parameters in the two models and are 0.2$\leq\alpha\leq$1.15 for model parameter $\alpha$, -0.79$\leq\tilde{\Pi}_{0}\leq$-0.33 for model parameter $\tilde{\Pi}_{0}$, 0$\leq\Omega_{m0}\leq$1 for the present value of the mass density parameter of dark matter $\Omega_{m0}$ and $H_0$ in the range 65-75 km $s^{-1}Mpc^{-1}$ by considering previous references and constraints imposed by various cosmological parameters.
The marginal likelihood function of $H_0$, $\alpha$ and $\tilde{\Pi_{0}}$ are plotted against respective parameter to study the distribution of those parameters specifically. We obtained true Gaussian curves centred almost at best estimated values of each parameters. The curves are fitted with Pantheon data of 1048 data points

The ratio of the likelihood of any two models gives the Bayes inference factor. We have obtained the Bayes factor of the viscous model in comparison with the standard $\Lambda$CDM model as 0.431. As per the Jeffreys criterion, this shows that the evidence of CVM model is not strong enough to be comparable with the standard $\Lambda$CDM model. Hence we can conclude that the standard $\Lambda$CDM model is found to have very strong evidence over the viscous model based on full Israel-Stewart theory, in predicting the late acceleration of the universe. 

\textbf{Acknowledgements} One of the authors VMS is thankful to CSIR for financial support through CSIR SRF fellowship.

\end{document}